\documentclass[finalversion, simpleeqnnos]{ylr-pdflatex}      % Specifies the ylr document class

\usepackage{graphicx,psfig,epsfig}
\usepackage{url}
\usepackage{amsfonts}
\usepackage{float}
\usepackage{multirow}

\ifx\pdfannotlink\dummy   % is the caller 'latex' or 'pdflatex'?
\usepackage[hypertex]{hyperref} % hyperlinks in xdvi?
\else
\usepackage[pdftex]{hyperref}   % generate pdf with hyperlinks
\fi

\floatstyle{ruled}
\newfloat{algorithm}{htbp}{loa}
\floatname{algorithm}{Algorithm}

\usepackage{latexsym,amsmath,amssymb}
\usepackage{graphicx}
\usepackage{tabularx}

\begin{document}
\title{Inventory Allocation for Online Graphical Display Advertising}

\author{
Jian Yang \hspace{0.2in} \hspace{0.2in} Erik Vee \hspace{0.2in} Sergei Vassilvitskii \hspace{0.2in} John Tomlin\\Jayavel Shanmugasundaram \hspace{0.2in} Tasos Anastasakos \\
Yahoo! Labs \\
701 First Ave\\
Sunnyvale, CA 94089\\
\{ {\tt jianyang, erikvee, sergei, tomlin, jaishan, tasos\}@yahoo-inc.com} \\
\vskip 10pt
Oliver Kennedy\\
Computer Science Department \\
Cornell University \\
Ithaca, NY 14840\\
{\tt okennedy@cs.cornell.edu}
}

\date{August 19, 2010}      % Deleting this command produces today's date.

\ylrnumber{YL-2010-004}   % The ASSIGNED report number

\newcommand{\ip}[2]{(#1, #2)}

%\numberofauthors{3}
%\author{
%  \alignauthor{Tasos Anastasakos} \\
%  \affaddr{Yahoo! Labs} \\
%  \affaddr{701 First Ave} \\
%  \affaddr{Sunnyvale, CA 94089.}
%  \email{tasos@yahoo-inc.com}
%  \alignauthor{Oliver Kennedy} \\
%  \affaddr{Computer Science Department} \\
%  \affaddr{Cornell University} \\
%  \affaddr{Ithaca, NY 14840.}
%  \email{okennedy@yahoo-inc.com}
%  \alignauthor{Jayavel Shanmugasundaram} \\
%  \affaddr{Yahoo! Labs} \\
%  \affaddr{701 First Ave} \\
%  \affaddr{Sunnyvale, CA 94089.}
%  \email{jaishan@yahoo-inc.com}\\
%\and 
%  \alignauthor{John A. Tomlin} \\
%  \affaddr{Yahoo! Labs} \\
%  \affaddr{701 First Ave} \\
%  \affaddr{Sunnyvale, CA 94089.}
%  \email{tomlin@yahoo-inc.com}
%  \alignauthor{Sergei Vassilvitskii} \\
%  \affaddr{Yahoo! Labs} \\
%  \affaddr{111 West 40th Street} \\
%  \affaddr{New York, NY 10018}
%  \email{sergei@yahoo-inc.com}
%  \alignauthor{Erik Vee} \\
%  \affaddr{Yahoo! Labs} \\
%  \affaddr{701 First Ave} \\
%  \affaddr{Sunnyvale, CA 94089.}
%  \email{erikvee@yahoo-inc.com}
%\and
% \alignauthor{Jian Yang} \\
%  \affaddr{Yahoo! Labs} \\
%  \affaddr{701 First Ave} \\
%  \affaddr{Sunnyvale, CA 94089.}
%  \email{jianyang@yahoo-inc.com}
%}
\date{}
\maketitle
\begin{abstract}
%\begin{abstract} 
We discuss a multi-objective/goal programming model for the allocation
of inventory of graphical advertisements. The model considers two
types of campaigns: guaranteed delivery (GD), which are sold months in
advance, and non-guaranteed delivery (NGD), which are sold using
real-time auctions. We investigate various advertiser and publisher
objectives such as (a) revenue from the sale of impressions, clicks
and conversions, (b) future revenue from the sale of NGD inventory,
and (c) ``fairness'' of allocation. While the first two objectives are
monetary, the third is not. This combination of demand types and
objectives leads to potentially many variations of our model, which we
delineate and evaluate. Our experimental results, which are based on
optimization runs using real data sets, demonstrate the effectiveness
and flexibility of the proposed model.
%\end{abstract}
\end{abstract}
%\category{H.4.m}{Information Systems}{Miscellaneous}
%\terms{Algorithms, Experimentation}
\section{Introduction}
\label{sec:intro}

Online graphical display advertising is a form of online advertising
where advertisers can explicitly or implicitly target users visiting
Web pages, and show graphical (e.g., image, video) ads to those
users. For instance, a brokerage firm may wish to target Males from
California who visit a Finance web site in the month of November 2010,
and show an ad promoting its special offers to those
users. Similarly, a different advertiser may wish to automatically
target users who visit a Finance web site, specifically those who are
likely to click on their ad highlighting a lower mortgage rate.
%where the ``set of users who are likely to click'' is automatically
%determined by a machine learning
%model~\cite{richardson07:_predic_click}.  
Online graphical display advertising is a multi-billion dollar
industry that is related to, but distinct from, sponsored search
advertising~\cite{AGM}, where advertisers bid for keywords
entered by users on a search page, and from content match
advertising~\cite{broder07:_seman_approac_contex_adver}, where
advertisers bid for clicks and text-matching techniques (as opposed to
user targeting) are used to show contextually relevant text
advertisements on Web pages.

As with most forms of online advertising~\cite{BroderRecSys08}, one of
the central questions that arises in the context of online graphical
display advertising is that of {\em inventory allocation}, i.e.,
determining how to allocate supply/inventory (user visits) to demand
(advertiser campaigns) so as to optimize for various publisher and
advertiser objectives. However, even formulating the inventory
allocation problem for online graphical display advertising is quite
challenging, for two reasons.

First, the same inventory can be sold in two different forms: {\em
guaranteed delivery} and {\em non-guaranteed delivery}. In guaranteed
delivery, an advertiser can purchase a certain number of targeted user
visits from a publisher several months in advance, and the publisher
guarantees these visits and incurs penalties if the guarantees are not
met. For instance, an advertiser may wish to purchase 100 million user
visits by Males in California on a Sports web site during Superbowl
2011, and the Sports web site publisher will guarantee these user
visits even though the serving date is months away from the booking
date. On the other hand, in non-guaranteed delivery, advertisers can
bid in real-time in a spot market for user visits, and the highest
bidder obtains the right to show an ad to the user. For instance, if a
user visits a Finance web page, then there may be multiple advertisers
bidding for the ad slot on the page, and the highest bidder can show
an ad to the user. An interesting aspect is that the {\em same user
visits} are eligible for both guaranteed delivery and non-guaranteed
delivery. A typical use case is that some inventory
%(e.g., user visits to Yahoo!\ Movies pages during Super Bowl 2011) is
not fully allocated to guaranteed campaigns can be sold to
non-guaranteed campaigns. However, not all visits from this inventory
will fetch the same price in the spot-market. This leads to the first
question addressed by this work: {\em How does a publisher allocate
inventory to both guaranteed and non-guaranteed advertising campaigns,
while still ensuring that the guaranteed advertiser objectives are
met, and publisher revenue is maximized?}

Second, unlike sponsored search and content match advertising, where
the goals of advertisers are to obtain clicks/ conversions on ads, the
goals of advertisers in online graphical display can be quite
varied. At one end of the spectrum are brand advertisers (e.g., major
department stores), whose primary goal is to reach a large and diverse
audience and promote their brand, rather than immediate clicks or
purchases. At the other end of the spectrum are performance
advertisers (e.g., credit card companies), whose primary goal is to
obtain immediate online clicks and conversions. In the middle, there
are performance-brand advertisers (e.g., car companies), whose goal is
both to promote the brand, as well as to obtain immediate leads of
users who are in the market to buy a car. The varied goals of advertisers
also lead to multiple currencies by which graphical display
advertisements are bought: brand advertisers typically buy impressions
(expressed in CPM, or Cost Per Mille (1000 impressions)), while
performance advertisers typically pay per click (CPC or Cost Per
Click) or conversion (CPA, or Cost Per Action), while
brand-performance advertisers may use a combination of CPM and
CPC/CPA. Thus, the second question we address is: {\em How does a
publisher allocate inventory across diverse advertisers and payment
types so that advertiser and publisher objectives are met?}

\subsection{Contributions}

Given the aforementioned unique requirements for online graphical display
advertising, one of the main technical contributions of this paper is
an inventory allocation optimization model that can capture these
requirements. At a high-level, the proposed allocation model
represents forecasts of future inventory (user visits) and guaranteed
advertiser campaigns as nodes in a bipartite graph. Each edge of the
bipartite graph connects a user visit to an eligible guaranteed
advertiser campaign. In addition, %since bids by non-guaranteed
%campaigns are not typically known until the time of serving, 
each user visit is annotated with a forecast of the highest bid fetched on the non-guaranteed marketplace and
a forecast of the expected pay-out. Similarly, each edge of the graph is annotated with a
forecast of the click or conversion probability for the advertiser
campaign and the specific user visit. 

Given the previous model, the objectives for online graphical display
advertising are captured as follows. There are two parts to the
objective function: one that captures guaranteed campaigns, and the
other that captures non-guaranteed campaigns. The objective for the
latter is simply to maximize the revenue for the publisher, since
advertisers only bid for what the user visit is worth to them. The
objective for the guaranteed campaigns, on the other hand, is more
complex because %there are penalties for missing
%guarantees, and 
advertisers could be interested in brand awareness, or
performance, or both.  Furthermore, the publisher faces penalties for under-delivering -- that is displaying an advertisement to fewer users than agreed on. 

In our model, delivery guarantees are treated as feasibility
constraints. (If the instance is infeasible, we trim the demand to
find the feasible solution with the minimum under-delivery penalties.)
The objective for guaranteed contracts has two parts. The brand
awareness objective is captured in terms of
``representativeness'' (see \cite{GMPV}), which tries to maximize the reach
of the guaranteed campaign by uniformly distributing the contracts among the user
visits to the extent possible
\footnote{Reach can also be specified in terms of users as opposed to user visits (as above) by using user cookie information, but we do not elaborate on this extension here}. 
The performance objectives for guaranteed campaigns are captured as the expected pay-out, i.e., the probability of clicks and conversions, times the value of each click
and conversion. 

Consequently, the final allocation objective has three parts:
non-guaranteed revenue, guaranteed representativeness, and guaranteed
clicks/conversions.  While multi-objective programming has been a
standard technique for some time, previous optimization models for
online advertising (see e.g.~\cite{LNAKK, AA}) have used a
single objective function.  One of our major contributions is to use
the multi-objective optimization framework~\cite{SR} to model the
sometimes conflicting objectives in a rigorous way.

While the multi-objective optimization model described above captures
the various objectives, it also introduces a new set of challenges
both in terms of operability and in terms of computational
feasibility.  Specifically, with regard to operability, the question
that arises is: how do we trade-off between the various objectives
(such as representativeness and non-guaranteed revenue), which do not
even have the same units? With regard to computational feasibility,
the question that arises is: how do we solve a
multi-objective formulation efficiently over large volumes of data
(tens of billions of user visits per day and hundreds of thousands of
advertiser campaigns per year)? Another key technical contribution of
this paper is a method that enables operators of the system to
trade-off between multiple objectives based on the monetary unit of a
single objective. For instance, an operator can trade-off
representativeness in terms of the impact it has on non-guaranteed
revenue, which is expressed in monetary units. A significant advantage
of this method is that it also allows for an efficient solution, which
can be solved on a small sample of the original bipartite graph,
without significantly compromising accuracy.

We note that there are two other closely related topics that impact
inventory allocation: {\em pricing} and {\em ad serving}. While the
details of these methods are beyond the scope of this paper, the
proposed allocation method works with quite general pricing and ad
serving techniques. Specifically, in terms of pricing, we assume that
guaranteed campaigns are priced using some external method (which may
itself use forecasts of non-guaranteed bids to ensure appropriate
prices), and the inventory allocation model can work with the booked
prices without regards to how exactly the prices were computed.
Similarly, we assume that we have some forecast of non-guaranteed
bids, but do not make any assumptions on how the bids by themselves
are generated by individual campaigns. In terms of ad serving, the
allocation model produces both a primal and dual solution, the latter
of which can be stored compactly and interpreted by the ad server to
serve ads to user visits in an online, real-time manner (e.g., see
~\cite{DevanurHayes2009, VVS}).

We have implemented the proposed inventory allocation model and the
solution techniques, motivated by the context of an operational online
graphical display advertising system. Our results using real user
visits, guaranteed campaigns, non-guaranteed bids, and
click/conversion data, indicate that the proposed approach is both
versatile in capturing and trading off between various advertiser and
publisher objectives, as well as efficient to solve with high
accuracy.

In summary, the main contributions of this paper are:

\begin{itemize}\itemsep=0in
\item A formalization of the inventory allocation model and objectives
      for online graphical display advertising (Section~\ref{sec:prelim}).
\item A multi-objective optimization formulation and various solution
      techniques for optimizing the multiple objectives (Section~\ref{sec:mathmodel}).
\item An experimental evaluation of the proposed techniques using
      real data sets (Section~\ref{sec:exp}).
\end{itemize}

\vspace{0.1in}
\section{Model and Objectives}
\label{sec:prelim}

We begin by first defining some notation, and then motivating and formalizing
the various objectives in online graphical display advertising.

\subsection{Supply and Demand Model}

As mentioned earlier, the main goal of inventory allocation is to
match supply (user visits) and demand (advertising campaigns). We thus
begin by modeling user visits, advertising campaigns, and their
interaction.

User visits can be represented as attributes-value pairs, where the
attributes represent the properties of a user, the properties of the
page they visit, as well as the time stamp of the visit. 
An example user visit could be represented as: Gender = Male, AgeGroup
= 30-40, Interests = \{Sports, Finance\}, Location = California, ...,
PageCategory = \{Sports\}, ..., Day = 15 Jan 2011, Time = 12:35pm GMT.

A display advertising campaign targets a subset of user visits by
specifying a targeting predicate. For instance, an advertising
campaign that targets Males in California visiting Sports pages in the
month of Jan 2011 can be represented as: Gender $\in$ \{Male\}
$\wedge$ Location $\in$ \{California\} $\wedge$ PageCategory $\in$
\{Sports\} $\wedge$ Duration $\in$ [1 Jan 2011 - 31 Jan 2011]. A user
visit is said to be eligible for an advertising campaign if the
attribute-value pairs of the user visit satisfy the targeting
predicate of the advertising campaign. In the rest of this paper, we
will focus on just the eligibility relationship between user visits
and advertising campaigns, and not on the specific attribute-value
pairs or the targeting predicates.

Advertising campaigns can be of two types: guaranteed campaigns,
whereby the publisher guarantees a fixed number of user visits to an
advertiser in advance, and non-guaranteed campaigns, where advertisers
bid in real-time for user visits. Both guaranteed and non-guaranteed
campaigns can have one or more advertiser goals: to obtain user visits
(this is the only goal that is guaranteed in a guaranteed campaign),
to obtain clicks, and/or to obtain conversions. In
non-guaranteed campaigns, however, these objectives are
converted into a bid by a bidding agent and thus, for the purpose of
modeling, they can be represented as a bid for each user
visit. Guaranteed campaigns, on the other hand, need to be modeled in
more detail. Specifically, a guaranteed campaign has a user visit
goal (the guarantee), a penalty function that specifies the penalty to
be paid by the publisher if the guarantee is not met, and a value for
each click and/or conversion. A key aspect that enables yield
optimization for clicks and/or conversions is the probability of a
click and/or conversion given a user visit. 

Finally, for reasons of scale, it is usually not practical to work
with all future user visits for a duration of many months (many large
publishers have billions of impressions {\em per day}). Consequently,
the inventory allocation problem often has to be solved on a sample of
user visits, and thus each sampled user visit is annotated with a
sample weight.

The following notation summarizes the above discussion:

\begin{itemize}\itemsep=0in
\item $\mathcal{I}$: Set of user visits
\item $s_i$: Sample weight of user visit $i \in \mathcal{I}$
\item $r_i$: The payout for user visit
      $i \in \mathcal{I}$ by non-guaranteed campaigns
\item $\mathcal{J}$: Set of guaranteed campaigns
\item $d_j$: User visit goal for guaranteed campaign $j \in \mathcal{J}$
\item $P_j: \mathbb{N} \rightarrow \mathbb{R}$: Penalty function for guaranteed campaign
      $j \in \mathcal{J}$, which maps under-delivery (how much a  
      guarantee is missed) to a penalty.
\item $W^c_j$: The value of a click for a guaranteed campaign $j \in \mathcal{J}$
\item $W^a_j$: The value of a conversion (action) for a guaranteed campaign 
      $j \in \mathcal{J}$
\item $p^c_{ij}$: The probability that a user corresponding to user visit
      $i \in \mathcal{I}$ clicks on an ad corresponding to guaranteed campaign
      $j \in \mathcal{J}$ 
\item $p^a_{ij}$: The probability that a user corresponding to user visit
      $i \in \mathcal{I}$ converts on an ad corresponding to guaranteed campaign
      $j \in \mathcal{J}$ 
\item $B_j$: Set of user visits $\in \mathcal{I}$ that are eligible for 
      guaranteed campaign $j \in \mathcal{J}$.
\end{itemize}

\begin{figure}[t!]
\centering
\includegraphics[width=0.6\textwidth]{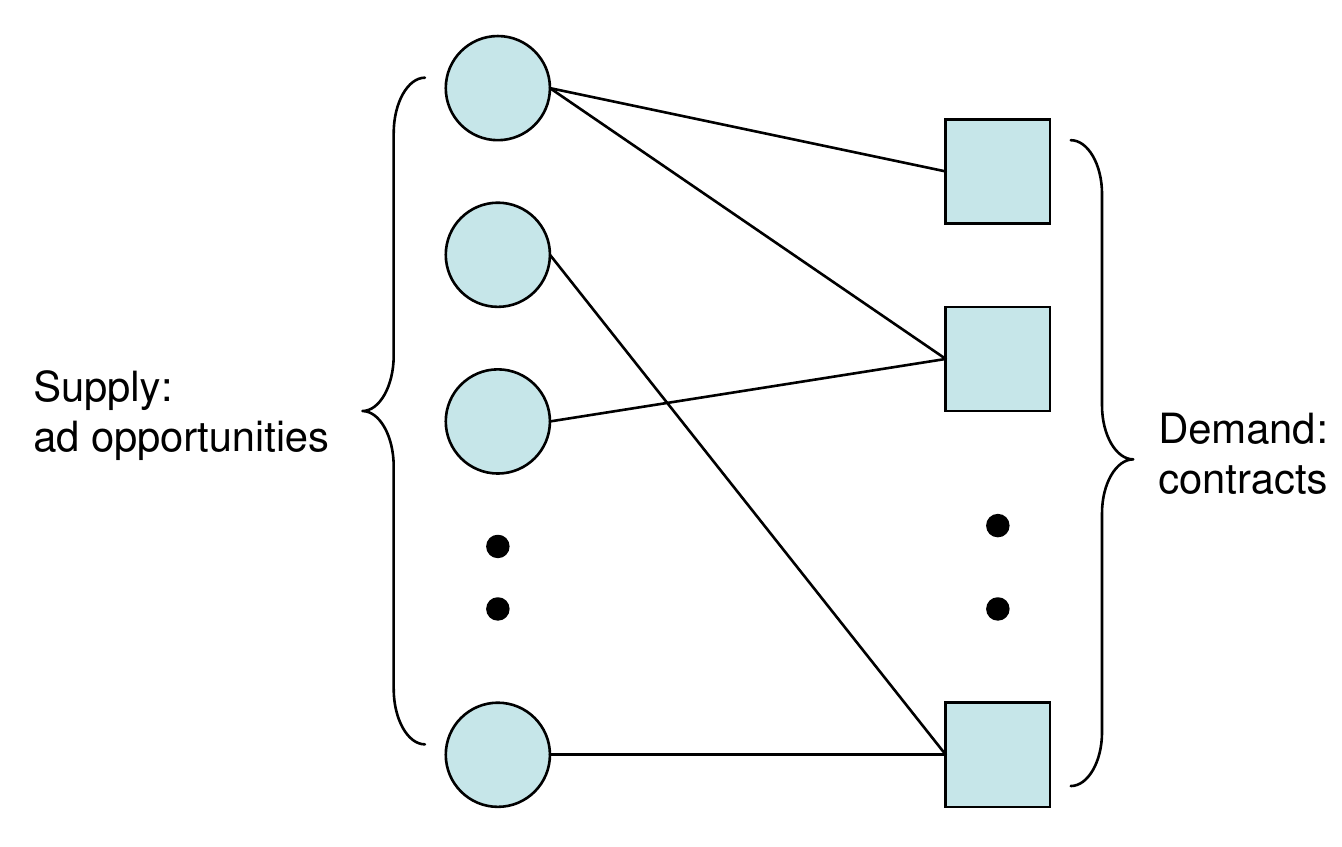}
\caption{Example of the allocation graph.}
\label{fig:graph}
\end{figure}

It is often convenient to view the user visits and guaranteed
campaigns in a bipartite graph, with user visits $\mathcal{I}$ on
one side, guaranteed campaigns, $\mathcal{J}$ on the other, and an
edge $(i,j)$ between a user visit $i$ and a guaranteed campaign $j$ if
$i \in B_j$, that is if $i$ satisfies the targeting predicates of $j$. An example of such graph is shown in Figure \ref{fig:graph}. 

It should be noted that the construction of this graph, which may involve tens of thousands of nodes and millions of edges, is in itself a major computational task.

\subsection{Objectives}

In this section, we motivate and formalize the various objectives that
are relevant to the inventory allocation problem. In the next section,
we introduce a mathematical model to trade-off and optimize across
these different objectives.

An overriding objective of publishers is to minimize the penalties
incurred in case guarantees are not met. Minimizing penalties is
important because the publisher not only incurs an immediate monetary
loss, but also because the publisher could suffer longer-term losses
due to advertiser attrition. Define $y_{ij}$ as the weight of user $i \in \mathcal{I}$ that is allocated to guaranteed campaign $j \in \mathcal{J}$. Then the amount of user visits delivered
to a guaranteed campaign $j$ is $\sum_i{y_{ij}}$, and the objective of
minimizing penalties can be formalized as:
\begin{equation}
    \min  \; \sum_j P_j(d_j - \sum_i{y_{ij}})
\end{equation}
However, in this paper we avoid this issue by ensuring feasibility of the constraints (see section \ref{make-feas}).

At a high-level, there are two parts to the inventory allocation
problem, corresponding to non-guaranteed and guaranteed campaigns.
The objective for non-guaranteed campaigns is simple: to
maximize the revenue for the publisher, since advertisers only bid
what each user visit is worth to them. Modeling the objectives for
guaranteed campaigns, on the other hand, is more complex because advertisers
could be interested in brand awareness/reach or performance (clicks,
conversions) or both. We thus have three objectives: non-guaranteed
revenue, brand awareness/reach, and performance.

Before formalizing the above objectives, we introduce some notation:
\begin{itemize}\itemsep=0in
\item $z_i = s_i-\sum_j{y_{ij}}$: The weight of user visit 
      $i \in \mathcal{I}$ that is allocated to non-guaranteed campaigns
\item $S_j = \sum_{i \in B_j} s_i$: The total amount of user visits
      eligible for guaranteed campaign $j \in \mathcal{J}$
\item $\theta_{ij} = s_i \frac{d_j}{S_j}$: The ideal fully representative
      target allocation fraction of user visit $i \in \mathcal{I}$ to
      guaranteed campaign $j \in \mathcal{J}$ (motivated further below)
\end{itemize}

\subsubsection{Non-Guaranteed Revenue}

The prices paid by non-guaranteed advertisers depend heavily on the particular user. Therefore, a natural goal of an allocation is to maximize the publisher's revenue by allocating the highest valued user visits to the non-guaranteed contracts.  Since the amount of
revenue that a publisher obtains from a user visit $i$ is $r_i z_i$,
this objective is written as:
\begin{equation}
    \max \; \sum_i r_i z_i
\end{equation}

\subsubsection{Brand Awareness/Reach}

There are two primary reasons why it is important to have a
brand awareness/reach objective. The first reason is that brand
advertisers typically want to reach a large swath of their target
audience. For instance, a brand advertiser who targets user visits from
the US will likely be quite unhappy if {\em all} of their delivered user
visits are from fourteen year old males in Wyoming, and none from the
rest of the population (even though all the delivered user visits
technically satisfy the targeting predicate of the advertiser). In
other words, brand advertisers often want a representative subset of
their target audience\footnote{Advertisers could potentially control their target allocation (other than uniformly at random) more closely if they desire, at the cost of targeting more defined, and therefore likely more expensive, inventory}.

The second reason for having a brand awareness and reach objective is more
subtle, and it relates to the interaction with non-guaranteed
campaigns. Specifically, if the primary goal of the publisher was to
maximize short-term revenue, then he could allocate all of the
high-value impressions to the highest bidding non-guaranteed
campaigns, and allocate only the remaining impressions to the
guaranteed campaigns. %(since the payments of the non-guaranteed
%campaigns are sold ahead of time, and the payment terms of those
%campaigns are only dependent on the number of user visits delivered,
%and not the ``quality'' of the user visits delivered). 
However, this is clearly detrimental to the advertiser, and is also a
dangerous road to take for the publisher: by selectively allocating the most
expensive user visits to the non-guaranteed
contracts, the publisher risks alienating in the long term the guaranteed advertisers,
many of whom pay a large premium for guarantees. In
fact as Ghosh et al.~\cite{GMPV} recently argued, in these situations
price serves as a signal of value, thus the user visits with the
highest $r_i$ may also be the ones most desired by the guaranteed
contracts.

Therefore, it is in the long-term interests of 
publishers to allocate each guaranteed campaign a {\em representative}
sub-set of targeted user visits. Ideally, every eligible user visit $i
\in B_j$ should be equally likely to see an ad from $j$. A similar
argument holds on a temporal scale. A week long contract should have
the same probability of being displayed on all days of the week.
That is, it should be allocated uniformly during the course of
the week---after all, if an advertiser wanted the contract to be shown
only on Tuesday, she would have added that to the targeting
constraints. 

To model these representativeness constraints, let $\theta_{ij}$
be the ideal target allocation. Note that
$\theta_{ij}$ directly encodes the fact that no user visit $i$ is
preferred by $j$ over others.  To maximize long term revenue, the
publisher should strive to find an allocation close to
$\theta_{ij}$. In this paper we use the $L_2$-norm distance to measure closeness to the target allocation. We denote $V_j$ the importance of a representative allocation to advertiser $j$ and write the objective as: 
\begin{equation}
\min \; \sum_j \sum_{i \in B_j}\frac{V_j}{2 \theta_{ij}}(y_{ij} - \theta_{ij})^2
\end{equation}
For consistency with other objectives, we will use the maximization form:
\begin{equation}
\max \; -\sum_j \sum_{i \in B_j}\frac{V_j}{2 \theta_{ij}}(y_{ij} - \theta_{ij})^2
\end{equation}

Note that alternative forms such as an entropy function or K-L
divergence can also be used, as they retain the essential features of
separability (by advertiser) and convexity. Similarly, other target
allocations besides the perfectly uniform target allocation can also
be considered but  these variants are beyond the scope of this
paper.

\subsubsection{Clicks/Conversions for Guaranteed Campaigns}

An advertiser in a guaranteed campaign may have multiple objectives,
such as clicks and conversions, besides obtaining the guaranteed user
visits. The publisher's objective is to maximize the yield from such
goals, while also directing such clicks and conversions to the
advertisers who value them the most. This can be modeled as trying to
maximize the expected value of clicks and conversions across all user
visits:
\[ \sum_j{\sum_{i \in B_j}{\left(W^c_j p^c_{ij} y_{ij} + W^a_j p^a_{ij} y_{ij}\right)}} \]
For compactness we define:
\begin{equation}\label{def-wij}
 w_{ij} = W^c_j p^c_{ij} + W^a_j p^a_{ij}.
\end{equation}
Then we can rewrite the performance objective as: 
\begin{equation}
\max \;  \sum_j \sum_{i\in B_j} w_{ij} y_{ij}. 
\end{equation}
Note that when we refer to ``clicks'' below, it is to be understood to include the subsequent conversions.

%\begin{align*}
%\mbox{Find } y_{ij} \mbox{ and } z_i & \mbox{ subject to: }\\
%   & \forall j\ \   \sum_{i \in B_j} y_{ij} = d_j  & \mbox{ (Demand Constraint)}\\
%           & \forall i\ \  \sum_{j | i \in B_j} y_{ij} + z_i = s_i &  \mbox{ (Supply Constraint)} \\
%         & \forall i,j\ \  y_{ij} \ge 0                        & \\
%         & \forall i\ \  z_i \ge 0                          & 
%\end{align*}

%\input{businessobjs.tex}
\section{The mathematical models}
\label{sec:mathmodel}

In the previous section we described the competing objectives faced by a publisher in allocating user visits to guaranteed contracts. In this section we formally state the mathematical problem that incorporates these objectives subject to the feasibility constraints.

There are three types of constraints that the allocation must satisfy to
be feasible. First, the desired number of user visits must be allocated
for each guaranteed contract:
\begin{eqnarray*}
\sum_{i \in B_j} y_{ij} = d_j & \text{(Demand Constraints)} & \ \ \ \ \forall j 
\end{eqnarray*}
Next, each user visit can be allocated to exactly one guaranteed or non-guaranteed contract.
\begin{eqnarray*}
  \sum_{j | i \in B_j} y_{ij} + z_i = s_i & \text{(Supply Constraints)} & \ \ \ \ \forall i
\end{eqnarray*}
Finally, we must ensure that the allocation is always non-negative.
\begin{eqnarray*}
y_{ij} \geq 0 &  \text{(Non-Negativity Constraints)} & \ \ \ \ \forall i,j\\
z_i \geq 0 & \forall i 
\end{eqnarray*}
Putting together the objectives from the previous section with the set of constraints, we may state our generic multi-objective optimization:
\begin{equation}\label{multi-obj}
    \max \; \begin{bmatrix}-\sum_j \sum_{i \in B_j}\frac{V_j}{2 \theta_{ij}}(y_{ij} - \theta_{ij})^2 \\
                 \sum_j\sum_{i \in B_j} w_{ij} y_{ij}\\ 
                 \sum_i r_i z_i \end{bmatrix}
\end{equation}
\vspace{-2mm}
\begin{eqnarray}\label{constraints}
    {\rm subject\ to}
               & \sum_{j | i \in B_j} y_{ij} + z_i = s_i & \forall i 
               \label{eqn:supplyConstraint}\\ 
    & \sum_{i \in B_j} y_{ij} = d_j & \forall j 
                \label{eqn:demandConstraint}\\
               & y_{ij} \ge 0                        & \forall i,j\\
               & z_i \ge 0                          & \forall i
	       \label{nonneg1}
\end{eqnarray}
%where the $w_{ij}$ are expected click value per impression as defined below in section \ref{single-obj}.

Note that the non-negative variables $z_i$ transform what would be
inequality supply constraints into equalities. Since $z_i$ is actually
the leftover supply inventory that will be sold to the non-guaranteed
delivery spot market, the term $\sum_i r_i z_i$ can be viewed as the
total non-guaranteed revenue that can be obtained for an allocation
$y$, with remnant inventory $z$. The second objective is the revenue
obtained from clicks on the displayed advertisements.

%We emphasize again that we have consistently cast our optimization problems as maximization problems. This requies no elaboration when we consider the revenue objectives. 

%--- now false after switch from min to max:  function, which is essentially a distance function. However, the possible conflicting revenue objective(s) are to be maximized, hence the appearance of their negatives in the multi-objective model. Thus when interpreting the efficient frontier figures in the following section the reader should keeep in mind that a larger value for ``representativeness'' actually corresponds to an increased, i.e. poorer, value of the distance functnonion, while increased revenue is of course desirable.

\subsection{Ensuring feasibility}\label{make-feas}

	      In this model formulation, guarantees are
 treated as constraints (Demand Constraints). Consequently, before we
optimize for the other objectives, we need to ensure that the model is
      feasible, i.e., there is sufficient supply for guaranteed
  campaigns. Note that even if a publisher is careful to accept only
  guaranteed campaigns that are feasible at the time of booking, the
 model could become infeasible at a later point because forecasts of
	  user visits could change due to unforeseen events.

In order to make the model feasible, we add dummy user visits that
have unlimited supply but a very high cost of being used, and connect
them to all the guaranteed campaigns. The cost associated with using
each dummy user visit for a guaranteed campaign $j$ corresponds to the
penalty $P_j > 0$ incurred by that guaranteed campaign in case of
under-delivery (note that if $P_j$ has multiple penalty values for
under-delivery, then these can be represented as multiple dummy user
visits, each with a different cost of being used). The cost associated
with using real user visits is 0. 

Given the above set up, the goal is to find the minimum cost
allocation to guaranteed campaigns. Since this problem is a pure
network with a linear objective function, it can be solved very
efficiently~\cite{BNO}. If the cost of the optimal allocation is 0,
then it implies that all the guaranteed campaigns can be satisfied,
and hence that the original model is feasible. If the cost of the
optimal allocation is greater than 0, then it implies that some
guaranteed campaigns will under-deliver. Furthermore, the optimal
allocation to the dummy user visits will indicate how much each
guaranteed campaign needs to under-deliver so that overall penalty
cost is minimized. In this case, the user visit goal $d_j$ for each
guaranteed campaign is reduced by the amount of allocation to the
dummy user visits in order to make the model feasible.  We then follow
one of the procedures described in the remainder of this section.

\subsection{Multi-objective programming}

\begin{figure}[ht]
{\centering
\includegraphics[scale=0.6]{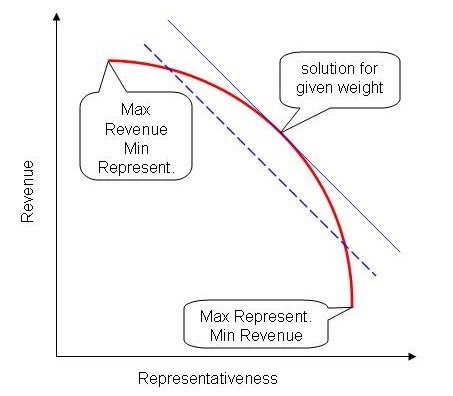}
\par}
\vspace*{-.1in}\caption{Efficient Frontier} \label{fig-1}
\end{figure}

We may approach a multi-objective function model in a number of ways (see~\cite{SR}). One general approach is to obtain an ``efficient frontier'' of solutions where at each point on the curve the value of one objective can only be improved at the expense of degrading another (see the generic example in Figure \ref{fig-1}, where we notionally trade off total revenue and ``representativeness''). The user may then choose any point on this curve as the ``solution''.

%\vskip 5pt
%\vskip 5pt

More algebraic approaches include using a weighted sum of the multiple objectives and/or using ``goal programming'', whereby the objectives are handled sequentially, with the additional constraint(s) that previous objectives retain a certain fraction of their optimal value. Since we have three objectives here there are clearly several variations, which we explore below.

In what follows, it will be convenient to define the 3 objective components as follows:

\begin{eqnarray}\itemsep=0in
F_1(y) &=& - \sum_j \sum_{i \in B_j}\frac{V_j}{2 \theta_{ij}}(y_{ij} - \theta_{ij})^2\\
F_2(y) &=& \sum_j\sum_{i\in B_j} w_{ij} y_{ij}\\
F_3(z) &=& \sum_i r_i z_i 
\end{eqnarray}

%Note that the signs are adjusted so that all the objectives are to be maximized.

Depending on the data available, we may formulate solution strategies which employ:
\begin{enumerate}\itemsep=0in
\item A single parametrized objective function.
\item A two-objective function model in various flavors
\item A three-objective function model, also in several flavors.
\end{enumerate}

\subsection{Single objective}\label{single-obj}

We consider a multi-component objective function:
\begin{equation}\label{F1F2F3}
\max \; \left\{ \gamma F_1(y) + \xi F_2(y) + F_3(z) \right\}
\end{equation}
subject to (\ref{eqn:supplyConstraint})-(\ref{nonneg1})   %(7)-(10).

The parameter $\gamma \ge 0$ is the weight for the ``representativeness'' component. The parameter $\xi$, in conjunction with the $w_{ij}$ reflects the means by which we attribute value to clicks. We observe that a feature of this model is that the shadow value $\beta_i$ of the supply constraint $i$ is always no less than the reserve price, i.e. $\beta_i \ge r_i$. %(We also note that $F_3(z)$ appears in the objective function of at least one of the sets of models described below).

Here, we assume that a value of $\gamma$ is available, from either historical data or business considerations. We consider explicitly two cases for the value of $\xi$:
\begin{itemize}\itemsep=0in
\item The values of the $w_{ij}$ are computed via (\ref{def-wij}) with $W_j^c = W_j^a = 1$. Then $\xi$ is the value of a click relative to a unit of revenue from the objective $F_3(z)$.
\item The $w_{ij}$ are actual expected monetary value of allocating an impression of type $i$ to contract $j$, in which case $\xi = 1$.
\end{itemize}
In both cases, the model involves the addition of a linear term in $y$ to the strictly convex distance function $F_1(y)$. The objective thus remains strictly convex quadratic in $y$ and linear in $z$. Optimization of this model is straightforward in principle, using a commercial solver such as XpressMP~\cite{xpressmp}. If a sufficiently powerful large-scale nonlinear network code were available, this could also be used since our constraints are of the pure network type.

\subsection{Two objectives}
\label{two-obj}

If all of the above data are not available to us, we must resort to multi-objective or goal programming. Let us first assume that $\gamma$ is not known, but the click-related data ($\xi, w_{ij})$ are available. In this case, we may initially solve:
\begin{equation}\label{f2f3}
\max \; \left\{ \xi F_2(y) + F_3(z) \right\}
\end{equation}
subject to (\ref{eqn:supplyConstraint})-(\ref{nonneg1}). 
Note that this is a linear pure network minimum
cost flow problem which can be solved very rapidly by special purpose
software (e.g., see Bertsekas~\cite{BNO}). Let the optimal value of
this linear program (LP) be $M^*$.

We may now append a constraint specifying that at least a certain fraction $\psi\,(0<\psi <1)$ of this monetary value be preserved and solve the model:
\begin{equation}\label{F1-min}
\max \; F_1(y)
\end{equation}
subject to (\ref{eqn:supplyConstraint})-(\ref{nonneg1}) and
\begin{equation}\label{M-LB}
\xi\sum_j\sum_{i\in B_j} w_{ij} y_{ij} + \sum_i r_i z_i  \ge \psi M^*
\end{equation}
It may be shown (see Appendix) that the unknown parameter $\gamma$ is equal to the inverse of the dual value for the constraint (\ref{M-LB}).

Suppose now that a value for  $\gamma$ is available but $\xi$ is not. We may initially solve
\begin{equation}
\max \; F_2(y)
\end{equation}
subject to (\ref{eqn:supplyConstraint})-(\ref{nonneg1}).
Let the optimal value of this linear program (LP) be $P^*$.
We may now append a constraint specifying that at least a certain fraction $\omega$ of this ``click value'' (however it is quantified) be preserved, and solve the model:
\begin{equation}
\max \; \left\{ \gamma F_1(y) + F_3(z) \right\}
\end{equation}
subject to (\ref{eqn:supplyConstraint})-(\ref{nonneg1}) and
\begin{equation}\label{click-LB}
\sum_j\sum_{i\in B_j} w_{ij} y_{ij} \ge \omega P^*
\end{equation}
The third variant combines the two objectives $F_1$ and $F_2$ in the second stage model, after solving a first stage model:
\begin{equation}\label{max-NGD}
\max \; F_3(y)
\end{equation}
subject to (\ref{eqn:supplyConstraint})-(\ref{nonneg1}). 
Since this is also a linear pure network minimum cost flow problem is can be solved very rapidly by special purpose software. Let the optimal value of this linear program be $R^*$

In this approach, the second model is of the form:
\[ \max \; \left\{ \gamma F_1(y) + \xi F_2(y) \right\} \]
\begin{equation}\label{ngd-LB}
\text{subject to (\ref{eqn:supplyConstraint})-(\ref{nonneg1}) and }
\sum_i r_i z_i  \ge \eta R^*
\end{equation}
where $\eta\,(0 < \eta <1)$ is the fraction of NGD revenue we wish to preserve.

This model would require us to have the relative weights on $F_1$ and $F_2$ available. This model is also guaranteed to be feasible if the original constraints (\ref{eqn:supplyConstraint})-(\ref{nonneg1}) are feasible.

\subsection{Three objectives}
\label{three-obj}

From an operability point of view, it is unlikely that both the
parameters $\gamma$ and $\xi$ would be known, or even understood very
well. In that case, we cannot avoid turning to goal programming and
the use of more intuitive ``knobs'' for the business to use in the
decision making process. Our goal program reduces to a sequence of
three models. In principle, these models could be solved in any order,
but we take the point of view that the nonlinear objective $F_1(y)$
should be optimized last, in order to avoid imposing a nonlinear
constraint. Such a non-linear constraint would be computationally
crippling at the scale we are contemplating. The last model therefore
must optimize representativeness subject to constraints on
non-guaranteed revenue and click value.

We again emphasize that, given appropriate $w_{ij}$, these last two
quantities can reasonably be thought of in monetary terms, whereas
$F_1$ may not.

Since non-guaranteed revenue is undeniably monetary, the most
intuitive procedure might be to first solve for $F_3$, i.e. solve the
problem (\ref{max-NGD}) and then solve for maximum click value:
\begin{equation}
\max \; F_2(y)
\end{equation}
\noindent subject to (\ref{eqn:supplyConstraint})-(\ref{nonneg1}) and
\begin{equation}
\sum_i r_i z_i  \ge \eta R^*
\end{equation}
where $\eta, R^*$ are as defined in constraint (\ref{ngd-LB}). Let the optimum objective function value be $P^{**}$. 

We may now optimize the representativeness function $F_1(y)$ subject to (\ref{eqn:supplyConstraint})-(\ref{nonneg1}) and
\begin{eqnarray}\label{double-LB}
\sum_i r_i z_i  \ge \eta R^* \\
\sum_j\sum_{i\in B_j} w_{ij} y_{ij} \ge \omega P^{**}
\end{eqnarray}

Clearly we could reverse the first two steps. This would likely
produce different results, but both would be feasible.

There is however a severe computational disadvantage to this 3 step
process. While step 1 is a pure network optimization model, the second
step involves a side constraint (\ref{ngd-LB}) or (\ref{click-LB}) which
destroys the orders of magnitude speed advantage of pure network
optimizers over general LP solvers. In an effort to avoid this, we
might omit the side constraint from the second model, leaving it a
pure network model. However there is now no guarantee that the third
model, incorporating (\ref{double-LB}), will be feasible. We might
then relax the side constraints until feasibility is obtained, but
rigor is lost.

\section{Experiments}
\label{sec:exp}
% Assigned to Jimmy (and Tasos?)
%% 5. Experiments
%%       5.0 Data set up and overview (Jimmy) 
%%         5.0.1 Contracts, steps to get to optimizer graph [Or whatever Jimmy wants]
%%       5.1 Forecast models
%%         5.1.1 SF Forecasting (Erik) 
%%         5.1.2 Forecasting (Sergei) 
%%         5.2.3 Click Forecasting (Tasos) 
%%       5.2 Data and Machine Description (Jimmy)
%%       5.3 Experimental Results (Jimmy (and Tasos??))
%%       5.4 Discussion (Jimmy (and (Tasos??))
%% -- Jimmy may restructure as he likes --

We now present some experimental results using a snapshot of real
online graphical display advertising data sets. The main goal is to
quantify the benefits of the multi-objective optimization approach
proposed in this paper as compared to the more traditional
single-objective optimization approaches. We note that our specific
focus is on quantifying the relative benefits of the various
optimization approaches for a given data/forecast snapshot, and not on
online ad serving mechanisms (e.g.,~\cite{DevanurHayes2009, VVS}) that
account for forecast errors, and could use periodic re-optimization.

\subsection{Experimental setup}

The data snapshot consists of guaranteed campaigns, forecasts of user
visits, non-guaranteed bids and clicks, all based on a subset of
historical data from an operational display advertising system. The
generated optimizer graph has 32,390 supply nodes, 2,696 demand nodes
and 1,407,753 edges.  The supply weight ($s_i$) ranges from $10.83$ to
$1.18\times10^9$.  The user visit goal ($d_j$) ranges from $1$ to
$6.96\times10^7$.  The non-guaranteed price ($r_i$) ranges from 0.046
to 4.350.  The click probability ($p^c_{ij}$) ranges from
$1.290\times10^{-6}$ to 0.947 (the experiments only evaluate clicks,
not conversions).

\begin{figure}[h!]
\centering
\includegraphics[width=0.6\textwidth]{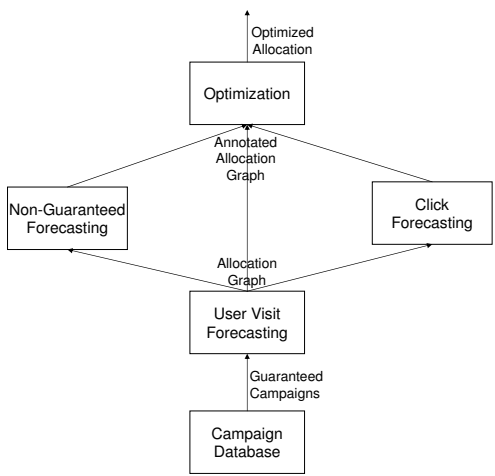}
\caption{Experiment Flowchart}
\label{fig:expSetup}
\vspace{-0.15in}
\end{figure}

Figure~\ref{fig:expSetup} shows the experimental flow based on the
data snapshot. The first step is to invoke the User Visit Forecasting
module to generate a forecast of user visits that correspond to the
guaranteed campaigns, and construct the allocation graph. The next two
(parallel) steps are to invoke the Non-Guaranteed Forecasting and
Click Forecasting modules to annotate the allocation graph with the
non-guaranteed revenue information ($r_i$) and the click probabilities
($p^c_{ij}$, respectively. The final step is to run the Optimizer to
solve the various optimization methods proposed in the paper.

The metrics that we measure for the various optimization runs are the
non-guaranteed revenue, the total value of clicks, and the
representativeness of guaranteed campaigns. The experiments are run
for various values of the model parameter settings. For the
experiments, all the allocation priorities $V_j$ are set to 1 and all the
click values $W^c_j$ are set to 10.

The experiments were run on a 64bit/64GB/2GHz Linux box using the
optimization package Xpressmp\footnote{Recently acquired by Fair Isaac (FICO)}~\cite{xpressmp}.

\subsection{Forecast models}
\label{sec:forecastModels}

Although the forecast models used here are not the focus of this
paper, we include a brief description of each for completeness.

\subsubsection{Supply Forecasting}

User visit forecasting uses time-series trend predictions to predict
the trend/growth in user visits for various pockets of supply such as
Sports or Finance. Our specific implementation uses
SARIMA~\cite{shumway07} for time-series predictions.

In addition, to generate the allocation graph, we need to produce a
sample of user visits. For this, we follow the sampling procedure
outlined in Vee et al~\cite{VVS}, which produces a unbiased sample for the class
of optimization models that we consider in this paper. Specifically,
for each guaranteed campaign, the supply forecast selects $k$ eligible
user visits uniformly at random. Once all of the user visits for all
campaigns are chosen, the weights given to each user visit is
normalized so that the total weight of any subset of user visits is
--- in expectation --- equal to the predicted available supply.  The
allocation graph is then created by adding edges from the campaigns to
their eligible user visits.

\subsubsection{Non-Guaranteed Forecasting}

The goal of the Non-Guaranteed Forecasting module is to predict the
expected revenue obtained by selling a particular user visit in the
non-guaranteed marketplace.  We observed that the prices paid for ad
opportunities by the non-guaranteed contracts followed a log normal
distribution, with prices ranging from under \$0.10 CPM to above
\$10 CPM. To predict the price of an individual user visit, we
trained a generalized least squares regression model on the logarithm
of the prices. Each user visit was annotated with a set of user
features, for example, age, gender, etc, and a set of page features,
for example Sports, Finance, etc. We used the value predicted by the
model as the non-guaranteed price $r_i$.

\subsubsection{Click forecasting}
\label{subsec:clickForecasting}

Click forecasting estimates the probability that a displayed ad will
be clicked in a particular user visit context.  Estimation of
click-through rates (CTR) is extensively applied in
pay-for-performance systems that attempt to maximize expected
revenue~\cite{richardson07:_predic_click,shaparenko09:_data_text_featur_spons_searc_click_predic}.
The estimates can be based on historical click-through performance
statistics of features that are selected as significant predictors of
CTR.  In this work, we use a logistic regression model with the
following functional form:
\begin{equation}
p(\text{click} \vert \{f_{1}, \ldots, f_{K}\}) = \frac{1}{1 +
  \exp(\sum_{k = 1}^{K}w_{k}f_{k})}
\end{equation}
where $f_{k}$ are predicates of features of the user visit and the ad
such as user age and gender, page content category, ad position in
page, etc.  Using historical data from web traffic that consist of
page visits with corresponding user visit and click statistics, we
learned the logistic regression parameters.

Finally, each edge in the allocation graph, which represents a
contract and a corresponding eligible ad opportunity, is annotated
with features $f_{k}$.  The click model is used to produce an estimate
of the probability of click $p_{ij}$ for each graph edge.

\subsection{Experimental results}

We now present our experimental results for various models and
parameter settings.  Each optimization for a particular parameter
combination takes about 5 to 10 minutes, which is quite acceptable as
an offline optimization step that can inform online serving algorithms
(e.g.,~\cite{DevanurHayes2009, VVS}).

\subsubsection{Baseline and Single-Objective Solutions}

As a baseline, we compute the optimization solution that only ensures
feasibility by minimizing the total under-delivery penalty, but does
not explicitly optimize for the other objectives. The result for this
baseline solution is shown in the first row of
Table~\ref{tbl:baseline}) and is used as the basis for normalization.
We then optimize for each single objective separately and report the
normalized solutions in the same table --- here NGD refers to
non-guaranteed revenue, and GD refers to guaranteed
representativeness.  It can be seen that while optimizing a single
objective may significantly improve that objective, it has a
potentially significant impact on the other objectives.

\begin{table}[h!]
\centering
\caption{Baseline and Single-Objective Solutions}
{\footnotesize
\begin{tabular}{|l|r|r|r|r|}
\hline
Objective   & NGD    & Click   & NGD+Click & GD \\
\hline
Baseline    & 1      & 1       &  1        &  -1\\
NGD         & 1.0165 & 1.0905  &  1.0229   &  -0.9836\\
Click       & 1.0062 & 2.9136  &  1.1722   &  -2.2214\\
NGD+Click   & 1.0131 & 2.9013  &  1.1774   &  -1.7440\\
GD          & 0.9968 & 0.9226  &  0.9903   &  -0.0027\\
\hline
\end{tabular}}
\label{tbl:baseline}
\end{table}

\subsubsection{Two-Objective Formulation}

\begin{figure}[h!]
\centering
\includegraphics[width=0.6\textwidth]{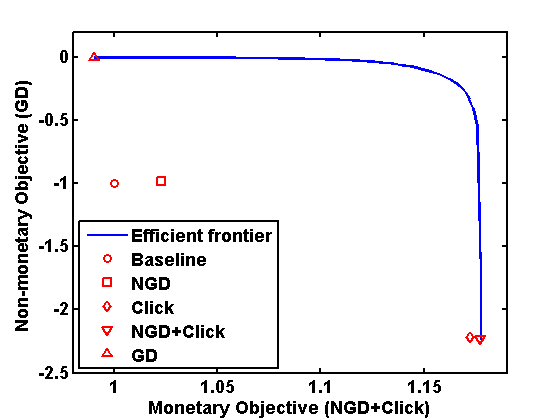}
\caption{Efficient Frontier}
\label{fig:2step}
%\vspace{-0.2in}
\end{figure}

We run the 2-step programming algorithm (Section~\ref{two-obj}) with
100 different values of the $\eta$ parameter.  Table~\ref{tbl:2step}
shows the objectives in normalized scale for a subset of the generated
efficient solutions.  The whole efficient frontier by using all the
100 points is depicted in Figure~\ref{fig:2step}, which shows the
trade-off between the monetary objective (non-guaranteed revenue plus
click value) and the non-monetary objective (representativeness).

\begin{table}[h!]
\centering
\caption{2-Step Goal Programming}
{\footnotesize
\begin{tabular}{|r|r|r|r|r|}
\hline
$\eta$  & NGD      & Click    & NGD+Click & GD \\
\hline
0.8411 &   0.9968 &   0.9226  &  0.9903 &   -0.0027\\
0.8570 &   0.9978 &   1.1269  &  1.0090 &   -0.0028\\
0.8729 &   0.9993 &   1.3260  &  1.0277 &   -0.0034\\
0.8888 &   1.0016 &   1.5171  &  1.0464 &   -0.0046\\
0.9047 &   1.0047 &   1.6990  &  1.0651 &   -0.0069\\
0.9205 &   1.0079 &   1.8810  &  1.0838 &   -0.0109\\
0.9364 &   1.0094 &   2.0801  &  1.1026 &   -0.0181\\
0.9523 &   1.0110 &   2.2782  &  1.1213 &   -0.0328\\
0.9682 &   1.0127 &   2.4763  &  1.1340 &   -0.0660\\
0.9841 &   1.0131 &   2.6861  &  1.1587 &   -0.1474\\
0.9999 &   1.0131 &   2.9013 &   1.1774 &   -1.7440\\
\hline
\end{tabular}}
\label{tbl:2step}
\end{table}

Figure~\ref{fig:2step} also shows the baseline and single-objective
solutions as different symbols.  The baseline solution and most of the
single-objective solutions are located in the interior of the feasible
solution region and dominated by efficient solutions on the
frontier. Further, the efficient frontier shows that a very small loss
in one objective can lead to a significant gain in the other
objective, thereby enabling a better combined trade-off as compared to
single-objective solutions.

The above results clearly demonstrate that explicitly capturing and
solving for multiple objectives dominates ignoring the objectives,
or just optimizing for a single objective.

\subsubsection{Three-Objective Formulation}

\begin{figure}[h!]
\centering
\includegraphics[width=0.6\textwidth]{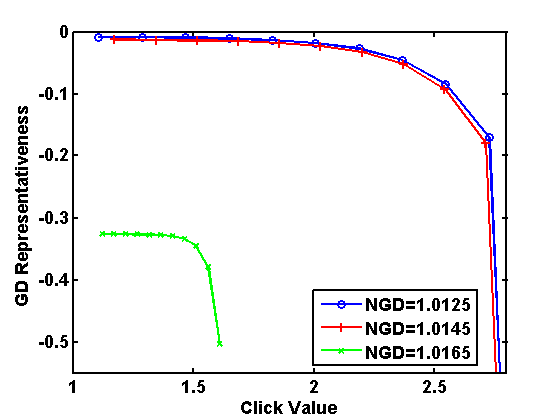}
\caption{Contours of Efficient Frontier}
\label{fig:3step}
\vspace{-0.2in}
\end{figure}

We run the 3-step programming algorithm (Section~\ref{three-obj}) with
100 different combinations of the $\eta$ and $\omega$ parameters.
Table~\ref{tbl:3step} shows the objectives in normalized scale for a
subset of the generated efficient solutions.  Note that the efficient
frontier for the tri-objective optimization problem is a surface in a
3D space.  One way to visualize the efficient frontier is to project
it to a 2D space by fixing the level in the third dimension.
Figure~\ref{fig:3step} shows three such contours in the space of
click value and GD representativeness for evenly-spaced NGD revenues.
The trade-off among the three objectives can be observed by comparing
points on the same contour and between different contours.

\begin{table}[h!]
\centering
\caption{3-Step Goal Programming}
{\footnotesize
\begin{tabular}{|r|r|r|r|r|r|}
\hline
$\eta$  & $\omega$ & NGD      & Click    & NGD+Click & GD \\
\hline
0.9961  &  0.3812 &   1.0125  &  1.1075  &  1.0208   & -0.0084\\
0.9961  &  0.6906 &   1.0125  &  2.0067  &  1.0990   & -0.0180\\
0.9961  &  1.0000 &   1.0125  &  2.9058  &  1.1772   & -1.7272\\
0.9981  &  0.4073 &   1.0145  &  1.1740  &  1.0284   & -0.0129\\
0.9981  &  0.7037 &   1.0145  &  2.0281  &  1.1027   & -0.0232\\
0.9981  &  1.0000 &   1.0145  &  2.8823  &  1.1770   & -1.7225\\
0.9999  &  0.6967 &   1.0165  &  1.1209  &  1.0255   & -0.3264\\
0.9999  &  0.8483 &   1.0165  &  1.3649  &  1.0468   & -0.3280\\
0.9999  &  0.9999 &   1.0165  &  1.6090  &  1.0680   & -0.5039\\
\hline
\end{tabular}}
\label{tbl:3step}
\end{table}

\subsection{Discussion}

The results above and especially the efficient frontier give us a
global picture of the trade-off among different objectives.  It
reveals what percentage of one objective can be gained at the cost of
one percent of another objective and how the trade-off rate changes
with the location of the solution, thereby providing valuable insight
in setting the right parameters in a real production
system to truly reflect the business priority. Perhaps more
importantly, it demonstrates the significant benefit of a combined
optimization model for multiple objectives, as opposed to optimizing
for just a single objective.

\section{Further extensions}

% Assigned to John and Oliver ???

We have referred throughout to {\em revenue} gained through clicks.
We might also include other non-monetary terms, using $\xi$ or some other parameter as a conversion factor. One candidate for this treatment is relevance, when $w_{ij}$ would encode the probability of relevance of ad $j$ to user $i$ and the payoff to advertiser $j$ for such relevance.

Another candidate for future work is the exploration of ways to to incorporte uncertainty into the strictly deterministic model we have described here. Preliminary stochastic programming experiments~\cite{BST2} indicate that uncertainty in the inventory supply can have a significant impact. Other sources of uncertainty arise in sampling error and click and conversion rate estimation.

\section{Related work}

Operations Research techniques, and optimization in particular, have
been used for decades in planning advertising campaigns in other
media, such as print, radio and television (see~\cite{BCPGSGH} for
example). The potential of the WWW for more targeted advertising was
realized in the 1990's and became a subject of research. Langheinrich et al~\cite{LNAKK} reported on a study where they attempted to target
display ads to users in order to optimize expected revenue, without
using intrusive data gathering techniques. While ``unintrusiveness''
seems to have become less of a concern to advertisers and publishers,
their optimization approach was influential. Tomlin~\cite{JT1}, noting
a similarity between this problem and the traffic distribution
model (see~\cite{AW1}), suggested adding an entropy term to the linear cost
function proposed in~\cite{LNAKK} to essentially smooth the
allocations and prevent ``bang-bang'' solutions; characteristic of LP
models. Chickering and Heckerman~\cite{352888} employed an alternate smoothing technnique, involving binning, and gave computational experience. A good survey of these and later developments is given in~\cite{AA}.

Since that time, the development of Computational Advertising as a
discipline has led to the consideration of many variants of the
graphical advertising allocation problem. A crucial development in
this process was the promulgation of the concept of ``fairness'' in
allocation to offset extreme LP solutions. While we have used the approach of~\cite{GMPV}, an alternate approach is taken 
by~\cite{DBLP:journals/corr/abs-1001-5076}. Some of our early work was presented in~\cite{ATY} and ~\cite{YT1}. Feldman et al~\cite{FMMM} have considered the 
online ad allocation problem in the more general setting of the matching
problem, while Roels and Fridgeirsdottir~\cite{RF} have considered a
dynamic version assuming the inventory follows a Markov
process. Easterly~\cite{AE1} and Han~\cite{Han1} discuss online
advertising from a Revenue Management point of view.

Two other modules in the online advertising supply chain make use of
the results of an allocation model such as we have described. These
are {\em Ad Serving} and {\em Admission Control}. The
Ad Server interprets the output of the allocation model as a set of
frequencies with which specific ads should be shown to users in the
supply pools when they visit a web page. Considerable practical
advantages ensue when the solution can be stored in compact form and
rapidly reconstructed on the fly by the ad server when a page requests
ads. This process has been studied by Devanur and Hayes~\cite{DevanurHayes2009} and Vee et al~\cite{VVS}, using the
dual values associated with the demand constraints and, implicitly,
the graph structure to determine the contracts for which a new
arriving impression is eligible. The optimization models we have
studied provide the necessary dual values, but online ad serving is
not considered in this paper.

Admission Control is the process of determining whether a proposed new
guaranteed campaign should be accepted, that is whether the existing
obligations can still be satisfied in a modified solution if the new
contract is accepted. Versions of this problem have been studied by
Feige et al~\cite{FIMN} and Aleai et al~\cite{AAKMMT} and considered
as an (NP-hard) combinatorial optimization problem. Radovanovic and Zeevi~\cite{RADZ} have proposed a relaxed, more tractable, variant of
the Admission Control process.

\section{Conclusion}

% Assigned to Jai and John

In this paper we have shown that multi-objective optimization provides
an efficient and flexible framework for the allocation of user visits
to guaranteed and non-guaranteed campaigns. The proposed models also
incorporate a very flexible means of taking into account revenue
derived from clicks, as well as non-monetary objectives --- in
particular ``representativeness'' or ``fairness'' --- in the
allocation of eligible user visits to campaigns. We are able to do
this using off-the-shelf software, such as a commercial scale
optimization system (we used XpressMP) for large scale quadratic
programming, or specialized network optimization codes for some steps.

We believe that the proposed models can be extended to other
objectives of interest such as ad relevance. As part of future
research, we are exploring extensions to the model to include
stochastic elements, such as uncertainty in supply and demand, and
nonlinear pricing.

\section{acknowledgments}

We gratefully acknowledge the assistance of Sumanth Jagannath and Wenjing Ma in processing the data for our experiments, and Jianchang Mao for managing much of the click model development.

\bibliographystyle{abbrv}
\bibliography{invalloc}
%\vspace{-0.15in}
\section*{Appendix}

The goal programming model (\ref{F1-min}), (\ref{eqn:supplyConstraint})-(\ref{nonneg1}), (\ref{M-LB}) in section \ref{two-obj} is equivalent to finding a value of $\gamma$ which preserves a specified fraction of revenue, given a value of $\xi$. To see this, let us write the Lagrangian of the problem (\ref{F1F2F3}), (\ref{eqn:supplyConstraint})-(\ref{nonneg1}) as:
\begin{eqnarray*}
    L^{\gamma} (y, z, \alpha^{\gamma}, \beta^{\gamma}, \lambda^{\gamma}, \mu^{\gamma}) = \gamma f(y, \theta) -\xi\sum_j\sum_{i\in B_j} w_{ij}y_{ij}\\
    - \sum_i r_i z_i
    - \sum_j \alpha_j^{\gamma} (\sum_{i \in B_j} y_{ij} - d_j) \\
    + \sum_i \beta_i^{\gamma} (\sum_{j | i \in B_j} y_{ij} + z_i - s_i)
    - \sum_{ij} \lambda_{ij}^{\gamma} y_{ij} - \sum_i \mu_i^{\gamma} z_i
\end{eqnarray*}
where the $\alpha_j^\gamma, \beta_i^\gamma, \lambda_{ij}^\gamma$ and $\mu_{ij}^\gamma$ are the appropriate Lagrange multipliers on the equality and inequality constraints.

The KKT optimality conditions for this problem are:
\begin{eqnarray*}
    \frac{\partial L^\gamma}{\partial y_{ij}} = \gamma \frac{\partial f}{\partial y_{ij}} - \xi w_{ij} - \alpha_j^{\gamma} + \beta_i^{\gamma} - \lambda_{ij}^{\gamma} = 0 \\
    \frac{\partial L^\gamma}{\partial z_i} = -r_i + \beta_i^{\gamma} - \mu_i^{\gamma} = 0 \\
    \lambda_{ij}^{\gamma} = \gamma \frac{\partial f}{\partial y_{ij}} - \xi w_{ij} - \alpha_j^{\gamma} + \beta_i^{\gamma} \ge 0 \\
    \mu_i^{\gamma} = \beta_i^{\gamma} - r_i \ge 0 \\
    \lambda_{ij}^{\gamma} y_{ij} = 0,\ \mu_i^{\gamma} z_i = 0
\end{eqnarray*}

Now consider the Lagrangian of the problem (\ref{F1-min}), (\ref{eqn:supplyConstraint})-(\ref{nonneg1}), (\ref{M-LB}), introducing Lagrange multiplier $\rho$:
\begin{eqnarray*}
    L^{\eta} (y, z, \alpha^{\eta}, \beta^{\eta}, \lambda^{\eta}, \mu^{\eta}, \rho) = f(y, \theta)\\
    - \sum_j \alpha_j^{\eta} (\sum_{i \in B_j} y_{ij} - d_j) 
    + \sum_i \beta_i^{\eta} (\sum_{j | i \in B_i} y_{ij} + z_i - s_i)\\
    - \sum_{ij} \lambda_{ij}^{\eta} y_{ij} - \sum_i \mu_i^{\eta} z_i \\
    - \rho (\xi\sum_j\sum_{i\in B_j} w_{ij} y_{ij} + \sum_{i} r_i z_i - \psi M^*)
\end{eqnarray*}
The KKT optimality conditions are:
\begin{eqnarray*}
    \frac{\partial L^\eta}{\partial y_{ij}} = \frac{\partial f}{\partial y_{ij}} -\rho\xi w_{ij} - \alpha_j^{\eta} + \beta_i^{\eta} - \lambda_{ij}^{\eta} = 0 \\
    \frac{\partial L^\eta}{\partial z_i} = -\rho r_i + \beta_i^{\eta} - \mu_i^{\eta} = 0 \\
    \lambda_{ij}^{\eta} = \frac{\partial f}{\partial y_{ij}} - \rho\xi w_{ij} - \alpha_j^{\eta} + \beta_i^{\eta} \ge 0 \\
    \mu_i^{\eta} = \beta_i^{\eta} -\rho r_i \ge 0 \\
    \lambda_{ij}^{\eta} y_{ij} = 0,\ \mu_i^{\eta} z_i = 0 \\
    \rho (\xi\sum_j\sum_{i\in B_j}w_{ij} y_{ij} + \sum_{i} r_i z_i - \psi M^*) = 0
\end{eqnarray*}
Comparing terms in the Lagrangians and the two sets of KKT conditions we see that $\alpha_j^{\eta} = \rho \alpha_j^{\gamma}, \ 
\beta_i^{\eta} = \rho \beta_i^{\gamma}, \ 
\lambda_{ij}^{\eta} = \rho \lambda_{ij}^{\gamma}, \ 
\mu_i^{\eta} = \rho \mu_i^{\gamma}$. 
In particular, we see that $\gamma \rho = 1$. Thus we may obtain the parameter $\gamma$ as $1/\rho$ from the solution to the goal programming problem. % It is not difficult to see that a similar result may be obtained for the other variants we have studied.

\end{document}